\documentclass{PoS}
 
\usepackage{epsfig}

\newcommand{\promille}{%
  \relax\ifmmode\promillezeichen
        \else\leavevmode\(\mathsurround=0pt\promillezeichen\)\fi}
\newcommand{\promillezeichen}{%
  \kern-.05em%
  \raise.5ex\hbox{\the\scriptfont0 0}%
  \kern-.15em/\kern-.15em%
  \lower.25ex\hbox{\the\scriptfont0 00}}

\newcommand{\bqa}{\begin{eqnarray*}}
\newcommand{\eqa}{\end{eqnarray*}}
\newcommand{\bq}{\begin{equation}}
\newcommand{\eq}{\end{equation}}

\def\bruch#1#2{\frac{\displaystyle#1}{\displaystyle#2}} 
\def\gehtzu{\raise0.3ex\hbox{$\longrightarrow$\kern-1.90em\raise-1.1ex
            \hbox{\footnotesize $\tau \rightarrow \infty$}}}
\def\gehtzuz{\raise0.3ex\hbox{$\longrightarrow$\kern-1.90em\raise-1.1ex
            \hbox{\footnotesize $z \rightarrow \infty$}}}

\ifpdf 
\DeclareGraphicsExtensions{.pdf}   % Dateiendung für PDF-Bilder
\else
\DeclareGraphicsExtensions{.eps,.ps}   % Dateiendung für PDF-Bilder
\fi

\PoS{PoS(LAT2005)164}

\title{Meson correlation functions at high temperatures}

\ShortTitle{Meson correlation functions at high temperatures}
  
\author{\speaker{S. Wissel}\thanks{BI-TP 2005/35, BNL-NT-05/30}, E. Laermann, S. Shcheredin \\
  Universit\"at Bielefeld\\
  Fakult\"at f\"ur Physik\\
  Universit\"atsstrasse 25\\
  D-33615 Bielefeld, Germany\\
  E-mail:\\
  \email{wissel@physik.uni-bielefeld.de},\\
  \email{edwin@physik.uni-bielefeld.de},\\
  \email{shchered@physik.uni-bielefeld.de}}

\author{S. Datta, F. Karsch\\
  Brookhaven National Lab\\
  PO Box 5000\\
  Phys. Dept., Bldg. 510C\\
  Upton, NY 11973-5000\\
  USA\\  
  E-mail:\\
  \email{saumen@bnl.gov}\\
  \email{karsch@bnl.gov}}

\abstract{We present preliminary results for the correlation- and spectral
  functions of different meson channels on the lattice. The main focus lies on gaining control over cut-off
  as well as on the finite-volume effects. Extrapolations of screening masses
  above the deconfining temperature are guided by the result of the free
  ($T=\infty$) case on the lattice and in the continuum. We study
  the quenched non-perturbatively improved Wilson-clover fermion as
  well as the hypercube fermion action which might show less cut-off effects.} 

\FullConference{XXIIIrd International Symposium on Lattice Field Theory\\
                 25-30 July 2005\\
                 Trinity College, Dublin, Ireland}

\begin{document}

\section{Introduction}
An important goal of contemporary physics is to understand the nature of excitations
characterizing the structure of QCD at high temperatures. At large
temperatures, the running coupling constant $g(T)$ becomes small and therefore
one can expect to see a weakly interacting gas of quarks and gluons. However,
at temperatures up to a few times $T_c$ some bound states might survive the
deconfinement transition.  
%it is not clear what are the characteristic excitations just above the
%transition temperature.
%In this work we aim to test different fermion discretizations and analyze the
%spatial correlators to shed light on this issue. 

Due to the small temporal extension of the lattice, it is a difficult task to reliably extract
information on bound states from temporal correlators. Like most
investigations so far, we first have studied spatial correlation functions,
but have attempted to carry out the infinite volume as well as the continuum
limit. Second, we employed the Maximum Entropy Method (MEM) to extract information
from the temporal correlators, working with and comparing two different
fermion discretizations with different UV cut-off effects. 
     
\section{Simulation details}
The results presented here are based on quenched gauge field configurations
generated with the standard Wilson plaquette gauge action. The physical
(temperature) scale has been obtained using the interpolation formula
of~\cite{Edwards:1997xf} for the string tension.  In order to study finite
volume and lattice spacing effects, at all considered temperatures, we have
generated gauge field configurations at three different volumes and lattice
spacings.  {\tiny
  \begin{table}[ht]
    \begin{center}
      \begin{tabular}{|c|c|c|c|c|l|c|}
        \hline
        $T/T_c$ & $\beta$ & $N_\sigma^3\times N_\tau$ & $a$ [fm] & $N_{conf}$ &  $\kappa$ & $\kappa_c$ \\ \hline
        
${\color{red} 1.24}$ & ${\color{red} 6.205}$  & $(64|32|24|{\color{red} \normalsize 16})^3\times 8$ & 0.075 & $(46|61|85|89)$ & 0.13599 & 0.13580\\
        
        1.24 & 6.499 & $(48|36|24)^3\times 12$ & 0.049 &$ (81|81|78)$ & 0.13558 & 0.13558\\
        
        1.24 & 6.721 & $(64|48|32)^3\times 16$ & 0.038 &$ (46|96|81)$ & 0.13507 & 0.13522\\

        1.19 & 6.872 & $64^3\times 20$ & 0.031 & 120 & 0.13494 & 0.13499\\ \hline
        
        $\color{red} {1.49}$ & {\color{red} 6.338} & $(64|32|24|{\color{red} \normalsize 16})^3\times 8$ & 0.062 &$ (46|62|59|60)$ & 0.13581 & 0.13578\\
        
        1.49 & 6.640 & $(48|36|24)^3\times 12$ & 0.041 &$ (61|59|68)$ & 0.13536 & 0.13535\\
        
        ${\color{red} 1.49}$ & ${\color{red} 6.872}$ & $(64|48|{\color{red} 32})^3\times 16$ & 0.031 &$ (66|62|60)$ & 0.13495 & 0.13499\\
        
        1.46 & 7.192 & $64^3\times 24$ & 0.021 & 80 & 0.13440 & 0.13450\\ \hline

        2.98 & 6.872 & $(96|64|32|24|16)^3\times  8$ & 0.031 &$ (37|43|65|80|80)$ & 0.13494 & 0.13499\\
        
        2.91 & 7.192 & $(48|36|24)^3\times 12$ & 0.021 &$ (85|80|80)$ & 0.13440 & 0.13450 \\
        
        {\color{red} 2.98 }& {\color{red} 7.457} & $(64|{\color{red} \normalsize
          48}|{\color{red} \normalsize 32})^3\times 16$ & 0.015 &$ (80|80|60)$ &
        0.13390 & 0.13394\\ \hline
        
5.96 & 7.457 & $32^3\times  8$ & 0.015 & 70 & 0.13390 & 0.13394 \\ \hline
      \end{tabular}
      \caption{Simulation parameters for the underlying study. (Data of the hypercube
        action are marked in red)}
      \label{param}
    \end{center}
  \end{table}
}
\vspace{-0.5cm} 

We use two different types of fermions: non-perturbatively $\mathcal{O}(a)$
improved~\cite{Luscher:1996ug} Wilson-clover
fermions~\cite{Sheikholeslami:1985ij} and hypercube
fermions~\cite{Bietenholz:1996pf}. At temperatures sufficiently above $T_c$
one is able to perform the Wilson matrix inversions in the chiral limit
%($\kappa =\kappa_c$) 
since the zero modes are absent or very suppressed. We
therefore choose to work at or close to $\kappa_c$ values, determined and
interpolated at $T=0$ from~\cite{Luscher:1996ug}, see Tab.~\ref{param}.

An alternative approach is to use the fixed point perfect action, which is
free from any cut-off effects. The existence of such an action is a
consequence of Wilsons renormalization group
theory~\cite{Wilson:1974mb}. Perfect actions have been introduced to the lattice
community in~\cite{Hasenfratz:1993sp}. For free quarks the perfect action is
known analytically and can be obtained with a technique called ``blocking from the
continuum''~\cite{Bietenholz:1995nk,Bietenholz:1995cy}. In lattice simulations one
truncates the perfect action to a unit hypercube to render it
ultralocal. To gauge the action we employed a simple ansatz for the
interacting gauge fields called the hyperlinks. This construction is supplemented by fat links and an additional enhancement of the operator couplings
for the free perfect action~\cite{Bietenholz:2002ks}. These provide the hypercube operator with better
chiral properties which are controlled by the shape of the eigenvalue
spectrum.  The latter is adjusted to hit the chiral
limit for the quark masses.
 
\section{Screening masses above $T_c$}

\begin{figure} 
\includegraphics[width=4.7cm]{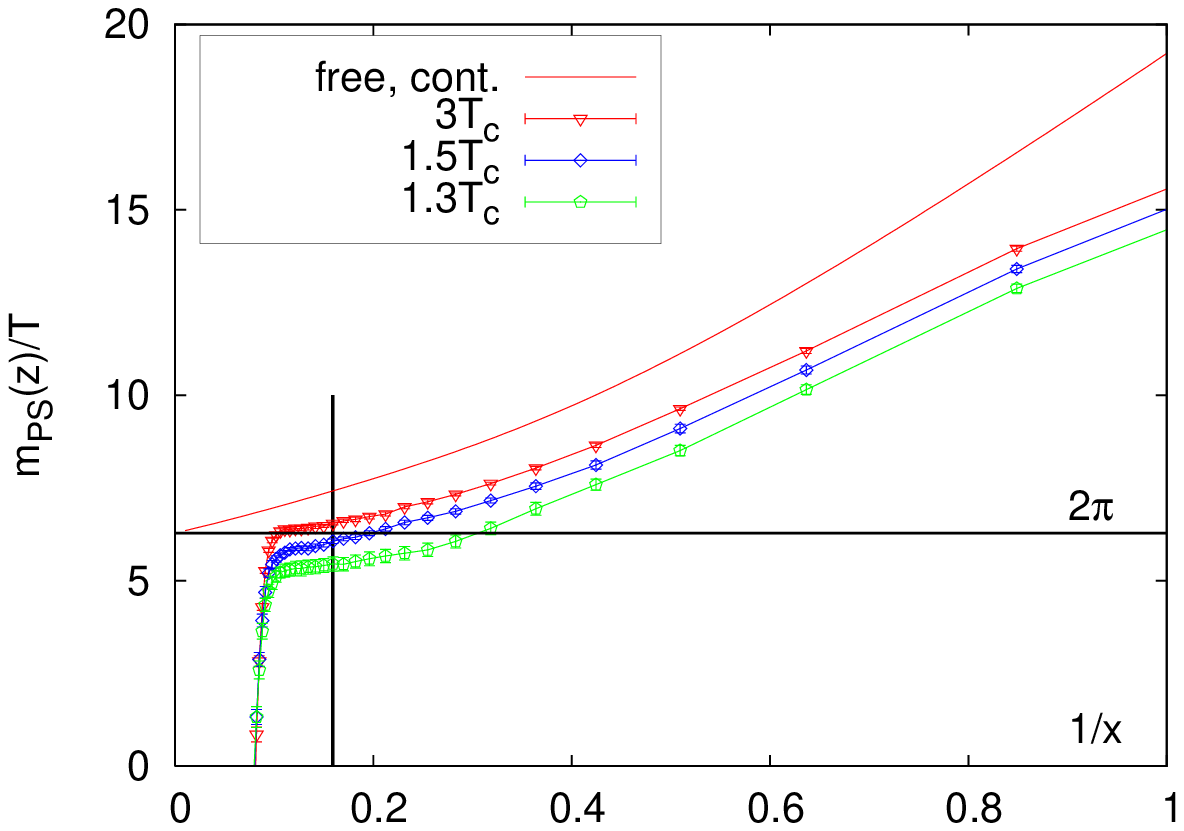}
\hspace{0.4cm}\includegraphics[width=4.5cm]{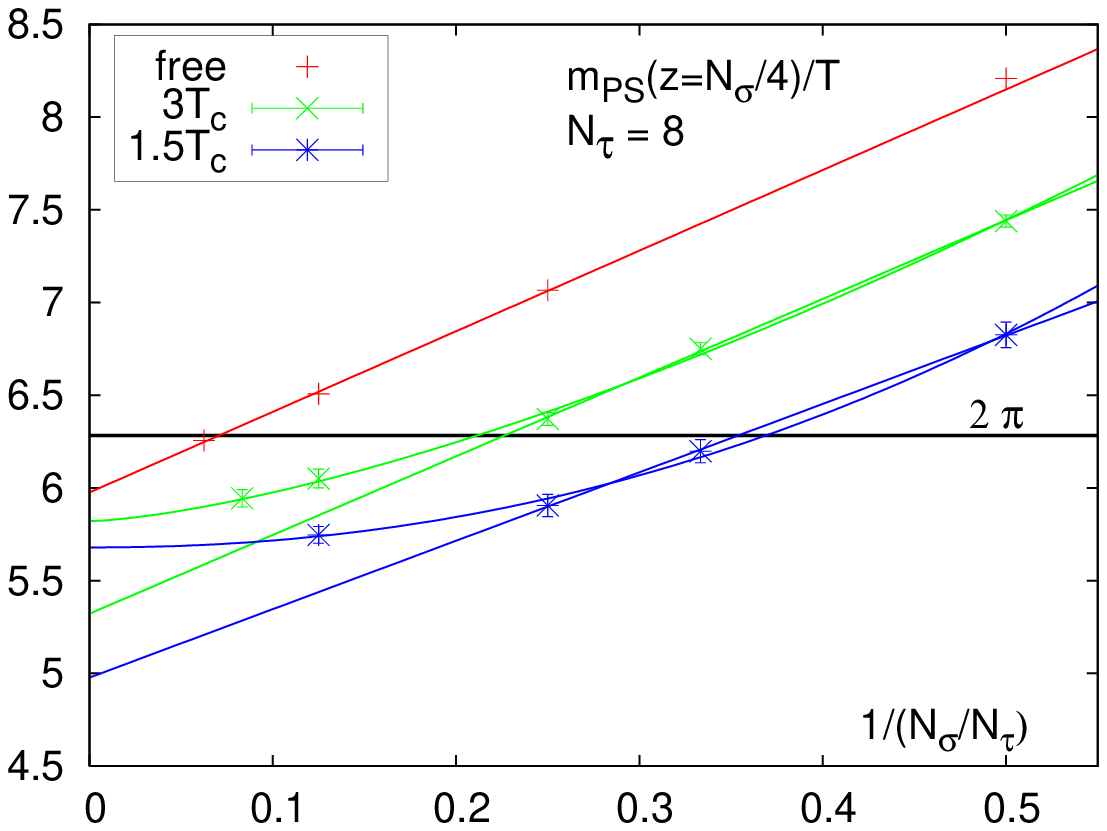}
\hspace{0.4cm}\includegraphics[width=4.5cm]{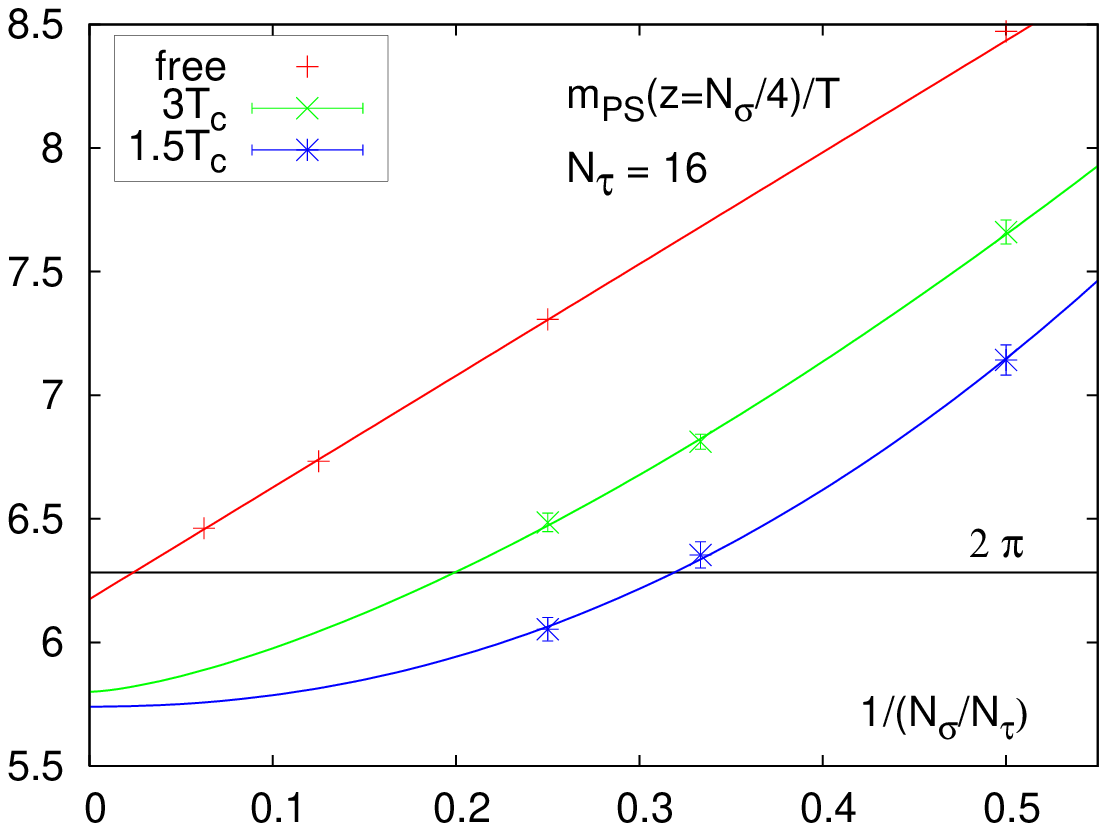} 
\caption{Effective screening masses compared to the free continuum one (left) and as function of aspect ratio (lattice volume)
  at $N_{\tau}=8,16$. The vertical line on the left plot corresponds to $z=L/4$.} 
\label{screen}
\end{figure}

In the infinite temperature limit mesonic correlation functions are
combinations of two free quark propagators. Correspondingly at large
distances the decay of spatial correlators is dominated by twice the lowest
quark Matsubara frequency,
\begin{equation} 
m^{\mathnormal meson}_{\mathnormal scr}=2\omega_0^{\mathnormal quark}=2\pi T.
\end{equation}
In the continuum the spatial correlation function of the free quark-antiquark
pair with pion quantum numbers is~\cite{Florkowski:1994rf} 
\begin{equation}  
G_{\pi}(z)=\bruch{N_cT}{2\pi z^2\sinh(2\pi Tz)}\left[1+2\pi Tz\coth(2\pi
  Tz)\right]\, .
\end{equation}
From this an effective z-dependent screening mass can be defined by
\begin{equation}
m(z)=-\bruch{1}{G(z)}\bruch{\partial G(z)}{\partial z}=2\pi T
\left\{\bruch{1}{x}
(2+x\coth(x))+\bruch{1}{1+x\coth(x)}\left(\bruch{x}{\sinh^2(x)}-\coth(x)\right)\right\}\, ,
\end{equation}
where $x=2\pi Tz$. 
%At large distances, $x\ge 5$ the exact data is reasonably
%well approximated by a leading behavior obtained as 
%\begin{equation}
%\bruch{m(z)}{T}=2\pi\left(1+\bruch{1}{x}+\ldots\right)
%\end{equation} 
Note that the effective screening mass, which is shown in Fig.~\ref{screen}
(left), does not exhibit any plateau-like behavior, which otherwise would
signal the presence of a genuine pole contribution. The anticipated value
$2\pi T$ is reached from above, yet only at asymptotically large
distances. Also in the interacting case a genuine plateau can not be
identified at temperatures above $T_c$ for our lattices. For definiteness we
therefore will quote screening masses obtained at a certain spatial
separation. The choice $z=L/4$ then leads in the free case to
\begin{equation}
\bruch{m(z=L/4)}{T}=2\pi \left( 1+\bruch{2}{\pi} \bruch{1}{LT}+\ldots
\right)
\end{equation} 
i.e. the leading correction to the continuum screening mass is linearly
dependent, $\sim 1/(LT)$, on the (inverse) separation.

In a numerical lattice investigation the separation is limited by the box
size.  In order to address finite volume effects we have calculated the free
spatial lattice correlation functions
semi-analytically~\cite{Karsch:2003wy}. It turns out, Fig. \ref{screen}, that
also in this case the leading finite volume correction to the screening mass
is linearly dependent on $1/(LT) = N_\tau/N_\sigma$. In the interacting case
the finite volume effects are less severe.  This can be understood because
freely propagating quarks feel the box boundaries (and the momentum cut-off)
more strongly than interacting quarks.  In order to describe the volume
effects we have therefore taken the exponent of the $1/(LT)$ term as a free
parameter,
\begin{equation}
\bruch{m(L,a)}{T}=\bruch{m(a)}{T}\left(1+\gamma\left(\bruch{N_{\tau}}{N_{\sigma}}\right)^p\right). \label{extravol}
\end{equation}
We expect that $p$ takes values between $3$ as in
the confined phase, e.g.~\cite{Fukugita:1992jj}, and $1$ as in the
infinite temperature limit. The fits
indeed return results which decrease towards $1$
with rising temperature, Table~\ref{tabone}.

\begin{figure}[ht] 
\begin{center}
\includegraphics[width=4.5cm]{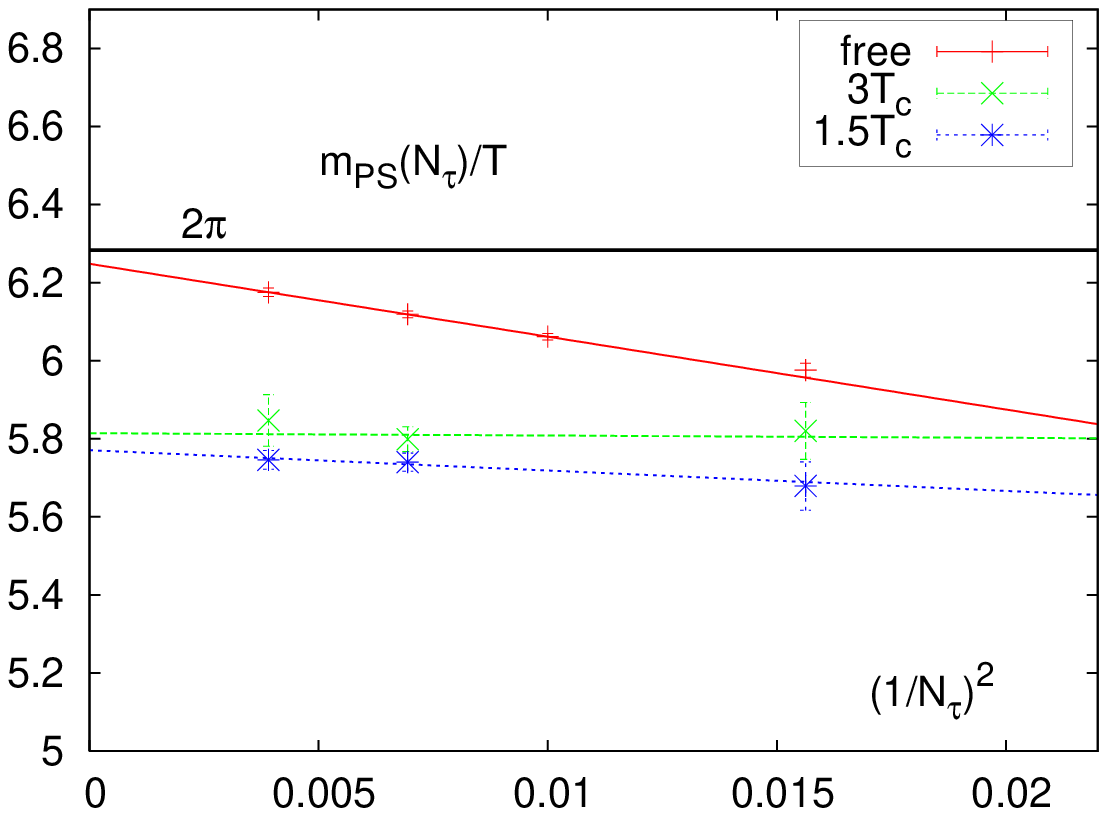}
%%%%%%%%%%%%%%%%%%%%%%%%%%%%%%%%%%%%
% This must be removed for pstopdf
\hspace{0.5cm}\includegraphics[width=7.8cm]{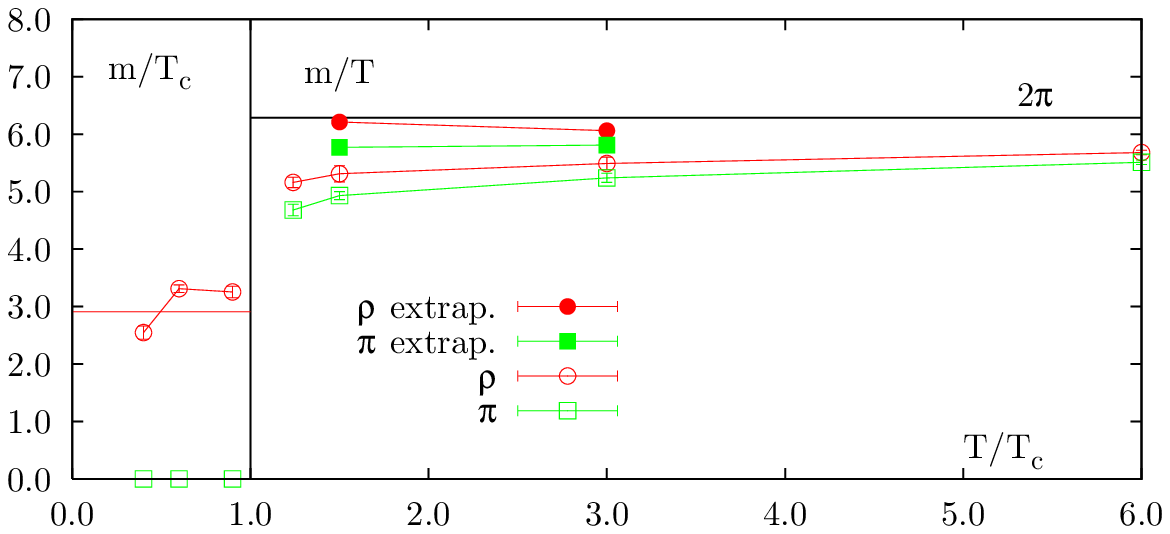}
%%%%%%%%%%%%%%%%%%%%%%%%%%%%%%%%%%%%
\end{center}%
\vspace{-3ex}
\caption{Finite lattice spacing extrapolation (left) and extrapolated
  result compared to old results, which have been obtained via rescaling with
  $2\pi / m_{\mathnormal free}^{\mathnormal scr}(L,a)$.}
\label{extra}
\end{figure}

The remaining lattice spacing dependence is expected
to be quadratic in $a$ or, equivalently in $1/N_\tau$,
\begin{equation}
\bruch{m(a)}{T}=\bruch{m_{\rm scr}}{T}-\lambda \left(\frac{1}{N_\tau}\right)^2+\ldots \label{extralat}
\end{equation}
in the free as well as in the interacting case.
As Fig. \ref{extra} (left) shows the coefficient $\lambda$
is rather small in the latter. The final numbers
are given in Table~\ref{tabone} and in Fig.~\ref{extra}.
\begin{table}[ht]
\begin{center}
\begin{tabular}{c|c|c}
& \multicolumn{2}{c}{$p$} \\ \cline{2-3}
   T & PS & V \\ \hline
%$1.2Tc$ & {\fr $3.57(60)$} & {\fr $2.71(50)$} \\
$1.5T_c$ & $2.12(36)$ & $2.24(46)$\\
$3T_c$ & $1.46(15)$ & $1.54(21)$\\
\end{tabular}
\hspace{1cm}
\begin{tabular}{c|c|c}
  & \multicolumn{2}{c}{$m_{\rm scr}/T$} \\ \cline{2-3}
   T & PS & V \\ \hline
%$1.2Tc$ & {\fr $6.04(6)$} & {\fr $6.13(15)$} \\
$1.5T_c$ & $5.77(1)$ & $6.21(1)$\\
$3T_c$ & $5.81(4)$ & $6.06(4)$\\
\end{tabular}
\vspace{-0.5cm} 
\end{center}
\caption{Results from the data fit to Eq.~(\protect\ref{extravol}) and ~(\protect\ref{extralat}) } 
\label{tabone}
\end{table} 
 
Due to exceptional configurations it is difficult to analyse the data below
$1.3T_c$.
%, which is not the case for $1.5T_c$ and $3T_c$.
As can be seen, the pion screening mass is about $9\%$ smaller than in the
free case while the vector meson comes out close to but also below $2 \pi$.
 This is to be contrasted with the analytical calculation based on dimensional
reduction and treating $\omega_0^{\mathnormal{quark}}$ as a heavy quark
mass~\cite{Laine:2003bd}, which predicts that the screening masses approach
the free result from above at $T\gg T_c$.  
%In the free case finite size and finite lattice
%spacing effects are found to be larger than in the interacting case. This is
%because freely propagating quarks are affected by the box boundaries and by
%the momentum cut-off more strongly than the interacting quarks.

\section{Spectral functions for the Wilson vs. the hypercube action}
\begin{figure}[ht]
%%%%%%%%%%%%%%%%%%%%%%%%%%%%%%%%%%%%%%%%%%
% This must be removed for pstopdf 
\includegraphics[width=4.9cm]{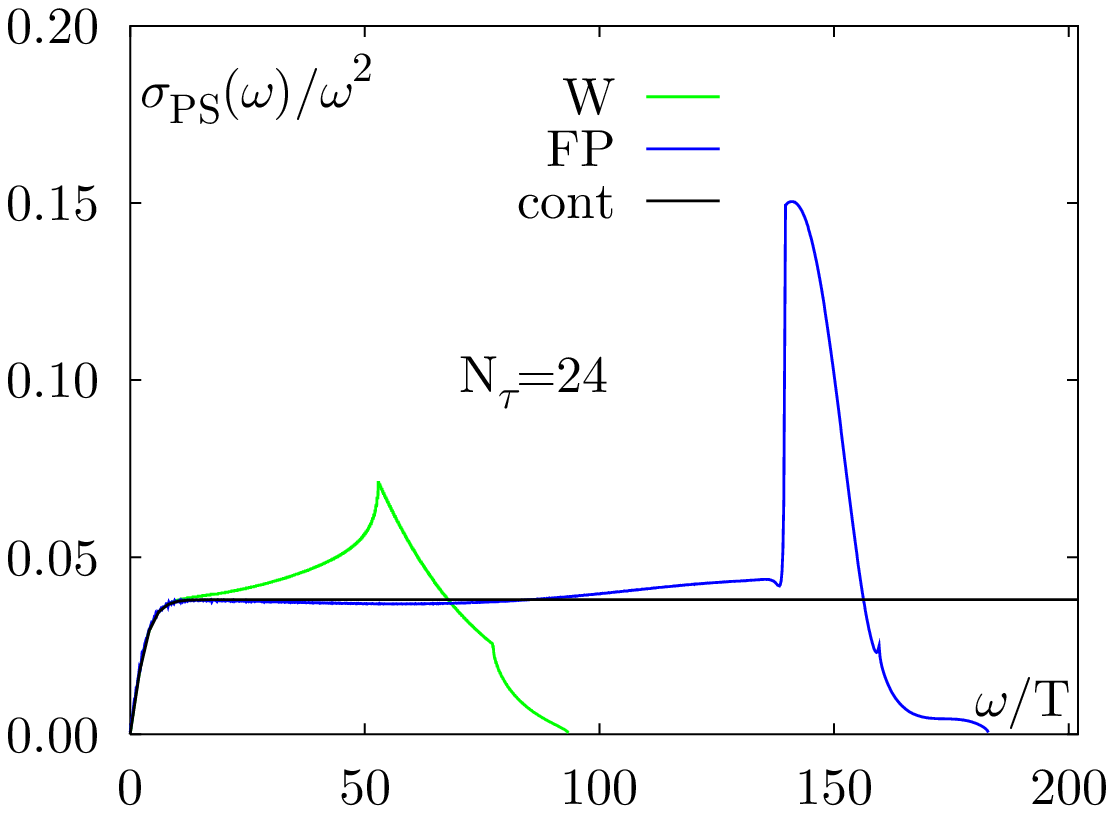}
%\includegraphics[width=4.9cm]{spec_fp_wf}
%%%%%%%%%%%%%%%%%%%%%%%%%%%%%%%%%%%%%%%%%%%
\includegraphics[width=5.0cm]{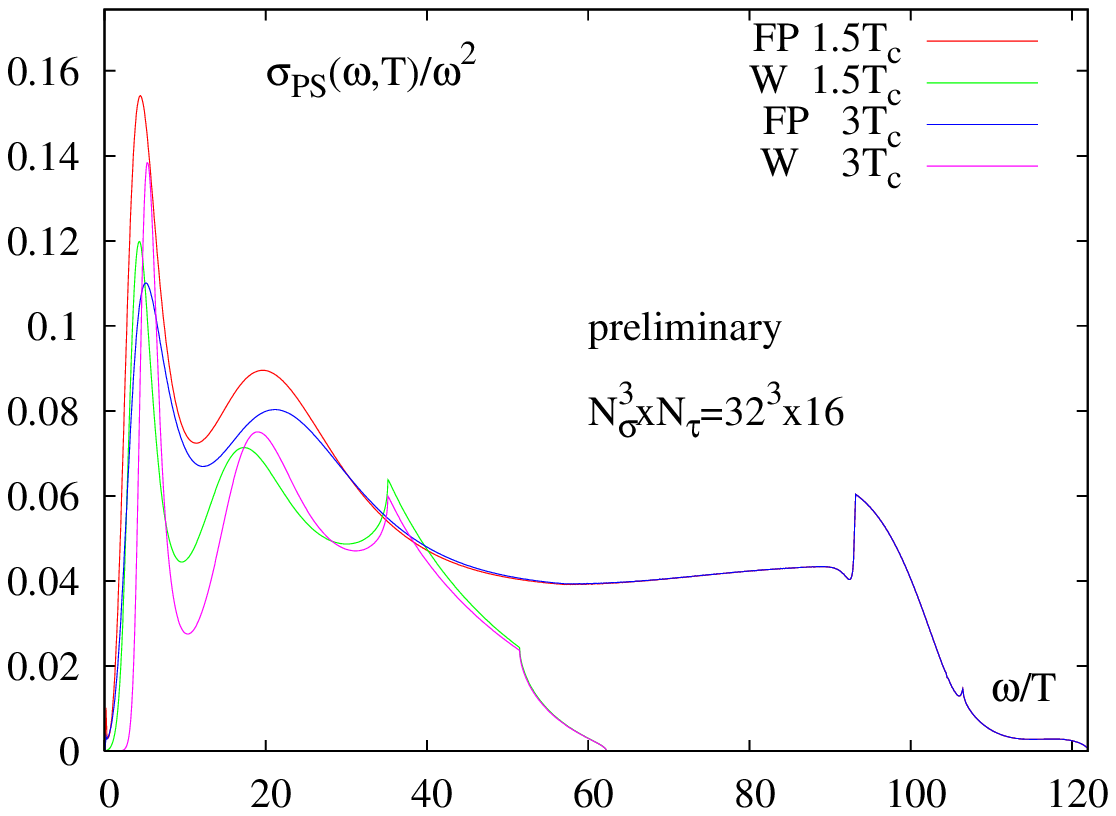}
\includegraphics[width=5.0cm]{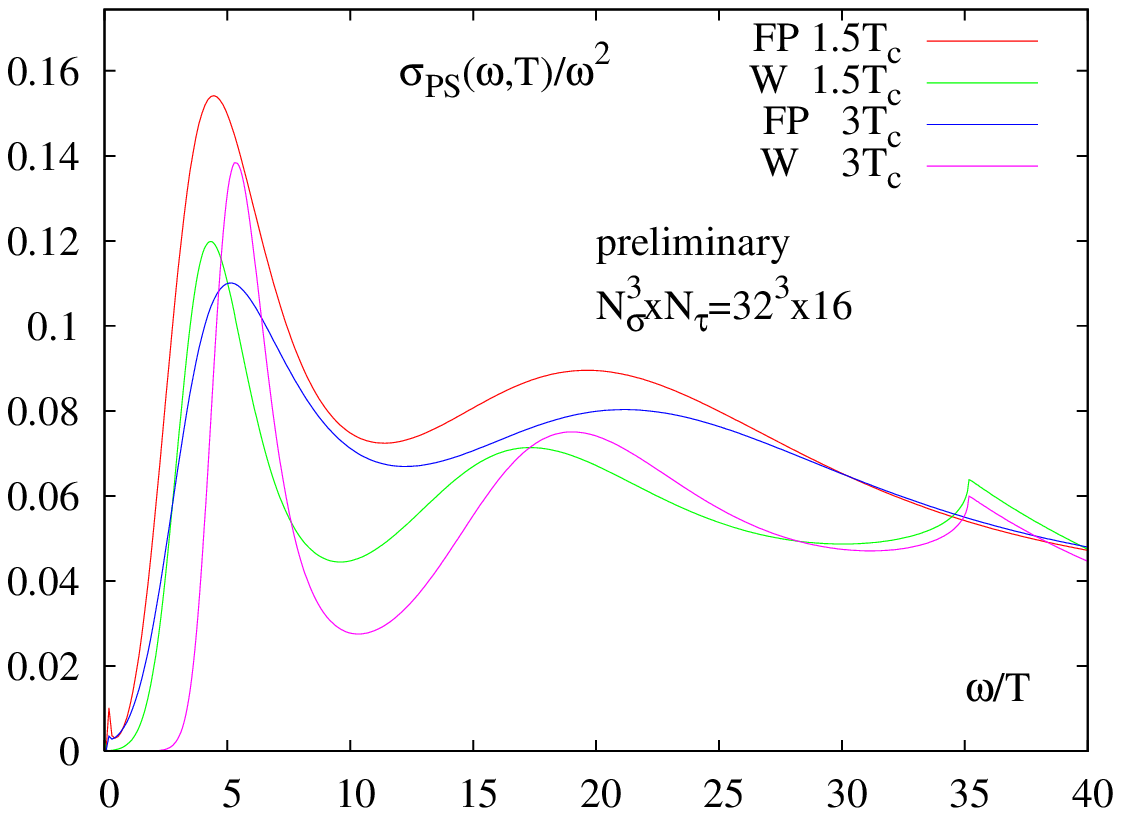} 
\caption{SPFs for the pseudo-scalar channel with Wilson (W) and hypercube
(FP) fermions at $T=\infty$ (left) and at $1.5T_c$, $3T_c$}
\label{FP_Spf}
\end{figure}

One important step towards the understanding of in-medium properties of mesons
 in finite temperature QCD is the study of their spectral functions
 (SPFs). They are necessary to establish connections with experimental results
 e.g. the dilepton or photon production rates and their mass spectrum. The
 computation of the SPFs on the lattice from the given correlation functions
 is an ill posed problem. However, one may attempt to use the well established
 MEM~\cite{Asakawa:2000tr}, which has been first utilized
 in~\cite{Nakahara:1999vy} to tackle the problem. In this method, information
 about the large $\omega$ part of the SPF, which is not well determined by the
 data, is supplied as a prior default in the form of a Shannon-Jaynes entropy
 term. The crucial point is to find a proper default model which serves as an
 input for MEM. To this end we consider the free lattice SPFs derived
 in~\cite{Karsch:2003wy,Aarts:2005hg,Aarts:2005zt} for our fermion actions.  In
 both cases MEM can nicely reproduce them with the input provided by the free
 correlation functions, see Fig.~\ref{FP_Spf} (left). In the free case the
 lattice artifacts for the hypercube fermions are pushed to much higher
 frequencies than for the Wilson fermions. The same behavior is seen also for
 the interacting case and thus allows for a more reliable study of interesting
 states in the physical interesting low $\omega$ region, see Fig.~\ref{FP_Spf}
 (middle). The reconstructed SPFs at $1.5T_c$ and $3T_c$ are significantly
 different from the free case. The positions of the first and the second bumps
 for both fermion discretizations are in agreement. The second bump in the
 case of the hypercube action corroborates, that it
may not be a lattice artifact, as one may have thought from studies with
Wilson fermions.  To solidify the SPF results from the hypercube action and to
turn them into predictions for the dilepton spectrum one needs renormalization
constants, whose determination is work in progress.
 
\section{Summary}
We have presented an analysis of infinite volume and continuum extrapolations
of the effective screening masses for different mesons with the improved
Wilson action above $T_c$. The resulting values come closer to the free
values.  SPFs computed with hypercube fermions seem to allow the
identification of  the excited states with higher confidence than Wilson fermions. To
fully confirm this behavior we further need to compute the renormalization
constants, improve our statistics and increase the volume.

\section*{Acknowledgments}
The computations have been performed on the IBM-JUMP computer at NIC
J\"ulich. This work was supported by the DFG under grant GRK 881. The work of
FK and SD has been partly supported by a contract DE-AC02-98CH1-886 with the
U.S. Department of Energy.
\bibliography{proclat05} \bibliographystyle{JHEP}

\end{document}